\documentclass[10pt,tightenlines,eqsecnum,floats,aps,showpacs,amsmath,amssymb,nofootinbib,superscriptaddress,prd,showkeys,twocolumn]{revtex4}

\usepackage{graphicx}

\usepackage{euscript,amssymb}

\usepackage{mathrsfs}

\usepackage{epsfig}
\usepackage{color}

\usepackage{amsfonts}
\usepackage{enumitem}
\usepackage{amsmath}

\usepackage{amssymb}

\usepackage{fancyhdr}

\usepackage{esint}
\usepackage[unicode=true, pdfusetitle,
 bookmarks=true,bookmarksnumbered=false,bookmarksopen=false,
 breaklinks=false,pdfborder={0 0 1},backref=false,colorlinks=false]
 {hyperref}

\begin{document}

\title{The final state of gravitational collapse in Eddington-inspired Born-Infeld  theory}

\author{Yaser Tavakoli}
\email{yaser.tavakoli@ut.ac.ir}
\affiliation{Department of Physics, University of Tehran,  North Kargar 14395-547 Tehran, Iran}
\affiliation{School of Physics, Institute for Research in Fundamental Sciences (IPM), 19395-5531 Tehran, Iran}
\affiliation{Departamento de F\'{\i}sica, Universidade Federal do Esp\'{\i}rito Santo, Av. Fernando Ferrari 514, 29075-910 Vit\'{o}ria - ES, Brazil}

\author{Celia Escamilla-Rivera}
\email{cescamilla@mctp.mx}
\affiliation{Mesoamerican Centre for Theoretical Physics, Universidad Aut\'onoma de Chiapas, Carretera Zapata Km. 4, Real del Bosque (Ter\'an), Tuxtla Guti\'errez 29040, Chiapas, M\'exico}
\affiliation{Departamento de F\'{\i}sica, Universidade Federal do Esp\'{\i}rito Santo, Av. Fernando Ferrari 514, 29075-910 Vit\'{o}ria - ES, Brazil}

\author{J\'ulio C. Fabris}
\email{fabris@pq.cnpq.br}
\affiliation{Departamento de F\'{\i}sica, Universidade Federal do Esp\'{\i}rito Santo, Av. Fernando Ferrari 514, 29075-910 Vit\'{o}ria - ES, Brazil}
\affiliation{National Research Nuclear University ``MEPhI", Kashirskoe sh. 31, Moscow 115409, Russia}

\begin{abstract}

In this paper, we address the implications when a homogeneous dust  model  is considered for
a scenario of  gravitational collapse in the context of Eddington-inspired Born-Infeld (EiBI) theory. 
In order to describe the dynamical evolution of the collapse, we present an effective equation, 
which constitutes the first order corrections, in EiBI coupling parameter $\kappa$, to Einstein's field equations. 
The geometry outside the collapsing object is derived by imposing the standard 
Darmois-Israel junction conditions at the boundary surface of the dust. 
This  induces  an effective matter source in the outer region   
which gives rise to a  non-singular, non-Schwarzschild  geometry at the final state of the collapse.
For  this  exterior geometry, we find  the threshold of mass for the  formation of the black hole.  This provides  a cut-off over $\kappa$ as $|\kappa|=5.1\times10^{-97} ~kg^{-1}\cdot m^3$. 

\end{abstract}

\date{\today}


\maketitle


\section{Introduction}
\label{intro}

General theory of relativity  is expected to be an incomplete theory due to the existence of singular solutions to the Einstein's field equations \cite{Hawking:1974,Joshi:2007,Wald:1984}. 
There are other frameworks for  gravity   which are free of space-time singularities.  One is the Einstein-Cartan theory of gravity with space-time torsion, which extends General Relativity (GR) to include quantum mechanical spin of elementary particles and resolves the space-time singularities 
\cite{Kibble:1961,Sciama:1964,Hehl:1976,Hehl:1974,Kuchowicz:1978,Pop lawski:2010}. 
It is also believed that the singularity problem will be overcome
in a quantum theory of gravity (see for example \cite{Bojowald:2007} and references therein). 
This issue has been widely studied in the context gravitational collapse (cf. see for example \cite{Bojowald:2005,Goswami:2006,Tavakoli:2013b,Tavakoli:2014} and references therein) and  cosmology \cite{Bojowald:2007,Ashtekar-Singh,APS:2006}.

An Eddington inspired Born-Infeld (EiBI) gravity theory was also proposed as an extension of Eddington's theory including  matter \cite{Deser,Banados:2010}. 
In this metric-affine proposal, the field equations are derived from a Lagrangian density where the variation is taken with respect to the both independent quantities, the metric $g_{ab}$ and the affine connection $\Gamma^{c}_{ab}$.
This effective modification allows  to remove the appearance of
cosmological singularities and is expected to be in agreement with GR at energies below the Planck scale. Nonetheless, the value of the EiBI coupling parameter $\kappa$ sets  a minimum length (or maximum density)  that points out to alternative scenarios where the singularities in gravitational collapse may be prevented. 
Recently, the astrophysical and cosmological issues of the EiBI gravity have been extensively
investigated 
\cite{EiBI-BH3,Scargill:2012kg,Pani:2012qd,Sotiriou:2010,Bouhmadi-Lopez:2013lha,Pani:2011mg,EiBI-BH1,EiBI-BH2,EiBI-BH4,EiBI-BH5,EiBI-BH6,EiBI-Perfect-Fluid}.  
It has been shown   that,  the avoidance of  the big bang singularity is not only limited 
to a radiation-dominated Universe. 
In Ref.~\cite{Pani:2012qd}  it was shown that the EiBI  theory shares some  pathologies with the $F(R)$ theory in the Palatini formalism \cite{Sotiriou:2010}, 
e.g., curvature singularities at the surface of polytropic stars,  and unacceptable Newtonian limit. 

Pathologies such as the surface singularities were shown to happen during the
phase transition inside a star \cite{Pani:2012qd}. In fact, since a star is made out of
elementary particles, this pathology problem may be cured when the
gravitational back-reaction on the matter dynamics is considered and the particles are effectively described by a polytropic fluid \cite{Kim:2013nna}, for which the
effective equation of state gets modified with the consequence that the surface are no longer singular. 
Other proposals to how remove such pathologies
include: to consider  a thick brane model \cite{Liu:2012rc} and by considering
matter sources with a time-dependent state parameter \cite{Avelino:2012ue}. These  ideas make
EiBI gravity a more consistent theory and a good prospective alternative
to GR.
In despite of that, it remains a very interesting theory to be applied in the high energy regime.
In the same context, it was found that,
the singularity is indeed present in the future Universe as a Big Rip where a phantom 
component is considered \cite{Bouhmadi-Lopez:2013lha}.

Since the EiBI gravity provides  a convenient  framework  to resolve the space-time singularities in  cosmological scenarios,
studying the  final  state of  the gravitational collapse  in this theory,  and the question how it may deviate from GR,  is  of  interest (e.g., see Ref. \cite{Pani:2011mg}). 
The Oppenheimer-Snyder model \cite{OS} is a simple scenario  for a   spherically symmetric, gravitational collapse 
of a homogeneous dust matter,  which provides a  convenient  analytical  framework  in order  to  study 
some  properties of a  gravitational system  in different  theories of gravity.
A particular result of this model is that, the final state of the dust collapse
will be a  (singular) Schwarzschild  black hole formation, which is indeed, a vacuum solution in GR.
In  this paper,  we are interested to study the Oppenheimer-Snyder-like model  in the framework of
an effective scenario of  EiBI  theory of gravity.

It is known  that in vacuum,  EiBI theory  is equivalent to GR, so that, 
the Schwarzschild metric is  a solution to the  EiBI action (see Eq.~(\ref{action-EiBI}) bellow) with no sources. 
However, in our herein model it will be shown that, when considering only up to the first order contributions of $\kappa$ in the EiBI equations of motion, we obtain an effective scenario of the collapse for which,  matching  the modified interior  region to a convenient  static    exterior space-time, through a  junction condition at the boundary, gives rise to an exterior  solution which is  different from the Schwarzschild geometry. The non-vacuum nature  of the exterior region  is a consequence of the fact that, the energy-momentum tensor of the effective field equation outside the collapsing dust  will contain an additional term induced by the modified interior region through the matching conditions. This analysis constitutes the main part of the  present work. Next, we will study  the geodesic behaviour of a massless particle propagating on the resulting exterior geometry and will present some of its interesting features.

\section{Eddington-inspired Born-Infeld theory}
\label{EiBI-theory}

EiBI theory is based on a Palatini formulation described by \cite{Banados:2010}
\begin{equation}
S = \frac{2}{\kappa}\int d^4x\Big(\sqrt{|g_{ab}+\kappa{\cal R}_{ab}|}-\lambda\sqrt{g}\Big) + S_{\rm matt}~,
\label{action-EiBI}
\end{equation}
where ${\cal R}_{ab}(\Gamma)$ represents the symmetric part of the Ricci tensor built from the  connection 
$\Gamma_{ab}^c$, and $\lambda$ is a dimensionless parameter which is related to the cosmological constant $\Lambda$ and the coupling parameter $\kappa$ through  
$\Lambda=(\lambda-1)/\kappa$; for  asymptotically  flat solutions  we have that $\lambda=1$. 
(This formulation is  given  in the Planck units, $8\pi G=1$.)
The matter action $S_{\rm matt}[g, \Gamma]$ is added in the usual way.
We observe also that,  in this formulation  a coupling parameter $\kappa$ is introduced, 
which is  a constant and  has  the    dimension   of  inverse $\Lambda$. 
When $\kappa {\cal R}_{ab}\ll 1$, the action
(\ref{action-EiBI}) reduces to the Einstein-Hilbert action with $\Lambda$. However, when
$\kappa {\cal R}_{ab}\gg 1$, the Eddington action is recovered \cite{Eddington-Book}:
\begin{equation}
S\ =\  2\kappa \int d^4x \sqrt{|{\cal R}|}\ .
\label{Eddington-pure}
\end{equation} 
In this case, by varying the  Eddington action (\ref{Eddington-pure}), integrating by parts and eliminating a vanishing trace, we obtain the field equations
\begin{equation}
2\kappa \sqrt{|{\cal R}|} {\cal R}^{ab} \ =\  \sqrt{|q|} q^{ab}.
\end{equation}
Hence, the parameter $\kappa$ 
interpolates between these  two different theories.

For the EiBI action (\ref{action-EiBI}), the equations of motion are given  by  the following complete set \cite{Banados:2010}:
\begin{eqnarray}
q_{ab}  &=& g_{ab}+\kappa{\cal R}_{ab}\ , \label{Eq-Motion1} \\
\sqrt{q}q^{ab} &=& \lambda\sqrt{g}g^{ab}-\kappa\sqrt{g}T^{ab}\ .  \label{Eq-Motion2}
\end{eqnarray}
Here,  $q^{ab}$ is the inverse of $q_{ab}$. 
Moreover, $T^{ab}$ is the standard energy momentum tensor whose  indices are lowered or raised with the metric $g_{ab}$ and its inverse.
Since  the action (\ref{action-EiBI}) reproduces Einstein gravity
within the vacuum, with $\Lambda$,
for a spherically symmetric configuration of this case, 
a Schwarzschild-de Sitter  metric  is,  thus,  a solution to the EiBI equation of motion with no sources \cite{Banados:2010}.
The corresponding  scalar curvature in this case,  is  ${\cal R}=4\Lambda=4(\lambda-1)/\kappa$.
In the presence of  matter in  EiBI  theory, two other aspects   have  been studied  in the   
regions with  high densities \cite{Banados:2010}:   a  black hole space-time  and  the very early
Universe,  in which  exact solutions can be seen
as minimum lengths that leads to singularity-free scenarios.

For the flat Friedmann-Lemaitre-Robertson-Walker  (FLRW) Universe, 
in the absence of cosmological constant  ($\Lambda=0$ or $\lambda=1$), 
the Friedmann equation is given by \cite{Banados:2010}
\begin{eqnarray}
H^2 =  \frac{1}{6}\frac{\mathbf{G}}{\mathbf{F}^2} \  ~,
 \label{Friedmann-formal}
 \end{eqnarray}
where $H=\dot{a}(t)/a(t)$ is the Hubble rate (a `dot' denotes a derivative with respect to $t$). Moreover,  $\mathbf{F}(\rho)$ and $\mathbf{G}(\rho)$ are defined as
\begin{eqnarray}
\mathbf{F}(\rho)  &:=&  
1- \frac{3\kappa(\rho+ p)(1-w-\kappa\rho-\kappa p)}{4(1+\kappa\rho)(1-\kappa p)}~, \\
\mathbf{G}(\rho)  &:=&  \frac{1}{\kappa}\left[1+2\sqrt{\frac{(1-\kappa p)^3}{1+\kappa\rho}}-3\frac{1-\kappa p}{1+\kappa\rho}\right] .
\label{main-eqn2}
\end{eqnarray}
In  the relations above, the standard matter, with the equation of state $p=w\rho$,  satisfies  the conservation equation $\dot{\rho}=-3H(\rho+p)$.
 Consequently, we can find the time derivative of the Hubble rate, $\dot{H}$,
 by using Eq.~(\ref{Friedmann-formal}) as
\begin{eqnarray}
2\dot{H}\ =\  H\left(\frac{\dot{\mathbf{G}}}{\mathbf{G}} - 2\frac{\dot{\mathbf{F}}}{\mathbf{F}}\right).
\label{Friedmann-der-formal}
\end{eqnarray}
Notice that, in equations above, we have worked in the units $8\pi G=c=1$.

\section{The modified dust collapse}

In the GR context, under variety of circumstances, singularities may appear at the final stages of gravitational collapse. Moreover, depending on  whether  trapped surfaces emerge early enough during the collapse, the singular region may be hidden behind a  {\em black hole} horizon.   Otherwise, the singular region will be visible to the distance observer, leading to  a  {\em naked singularity}  formation  as  the collapse outcome  (see Ref.~\cite{Joshi:2007} and references therein).

The problem of collapsing scalar fields cosmologies, in the context of GR, has been studied in the literatures (see for example \cite{Giambo:2005,Giambo:2008,Tavakoli:2013a}). Therein, solutions corresponding to the black hole and naked singularity were found as possible collapse end states.
For  a spherically symmetric homogeneous dust collapse in GR (i.e., an  Oppenheimer-Snyder model \cite{OS}), trapped surfaces  always do form at the boundary of the star, so that the final singularity will be covered by a Schwarzschild black hole horizon.

In what  follows, motivated from Refs.~\cite{Giambo:2005,Giambo:2008,Tavakoli:2013a,Goswami:2006,Tavakoli:2013b,Tavakoli:2014}, we consider  a flat FLRW geometry%
\footnote{In order to employ the  EiBI theory of flat FLRW cosmology, as we presented in previous section, we consider a flat FLRW geometry for our stellar collapse. Nevertheless, we expect that (similar to  GR context),  the emerging exterior metric in our model should not be different from that raised  by a closed FLRW interior (i.e. the Oppenheimer-Snyder model).}
for  the {\em interior}  space-time of a collapsing star whose matter source is given by a homogeneous dust cloud. The interior line element reads,
\begin{eqnarray}
ds^2_{\rm (int)}\ =\ -dt^2 + a^2(t)(dr^2 + r^2d\Omega^2)\ ,
\label{metric-interior}
\end{eqnarray}
where $d\Omega^2=d\theta^2+\sin^2\theta d\phi^2$.
We analyse the dynamical evolution of
the collapse  by employing an effective scenario of EiBI theory, we introduced above.
Then, we will study the possible exterior geometry of the collapse by a matching to the interior at the boundary of the cloud.

\subsection{The interior space-time dynamics}

For a dust matter ($w=0$, $p=0$), 
in  the interior  region  (\ref{metric-interior}),
we can expand  Eq.~(\ref{Friedmann-formal}) up to second order in $\kappa\rho$), to find  the first  order\footnote{For a discussion  
on    higher order contributions of  $\kappa$, see appendix \ref{appendix1}.} correction terms  to  the standard general relativistic Friedmann equation%
\footnote{The other way of deriving modified Friedmann equation is to  expand first the original field equations (\ref{Eq-Motion1}) and (\ref{Eq-Motion2}), to `second' order in $\kappa$ to find the `first' order corrections to the Einstein's equation \cite{Banados:2010}:
\begin{eqnarray}
\mathcal{R}_{ab}\ \approx \ \Lambda g_{ab} + T_{ab} - \frac{1}{2}Tg_{ab}+ \kappa\big(S_{ab}-\frac{1}{4}Sg_{ab}\big), 
\label{Einstein-Mod}
\end{eqnarray}
where $S_{ab}\equiv T_a^c T_{cb} -\frac{1}{2}TT_{ab}$.
Then, by replacing the flat FLRW geometry (\ref{metric-interior}) with the dust matter source
in the improved Einstein's equation (\ref{Einstein-Mod}), the `first' order modified Friedmann equation (\ref{Friedmann}) is obtained.}:
\begin{eqnarray}
3\frac{\dot{a}^2}{a^2}\    \  =\    \rho\Big(1+ \frac{3}{8}\kappa\rho\Big)  + {\cal O}\big(|\kappa|^2\big) \ =:\  \rho_{\rm eff} \  .
\label{Friedmann}
\end{eqnarray}
Here $\dot{a}<0$ indicating a collapse process.
Eq.~(\ref{Friedmann}),   represents  two different scenarios for the final state of gravitational collapse,
depending on the sign of $\kappa$:  
if $\kappa<0$,  the energy density $\rho$ of the collapsing cloud  starts  increasing from the initial density $\rho_0=\rho(0)$,
until it  reaches a maximum  $\rho_{\rm max}=8/3|\kappa|$ at  which the Hubble rate (\ref{Friedmann}) vanishes.
Therefore,  the general relativistic singularity  is resolved, in this case,  and is replaced by a  bounce. 
Notice that, the effective density is always positive, $\rho_{\rm eff}\geq0$, 
and the matter density  changes in the interval  $\rho_0\leq\rho\leq \rho_{\rm max}$.
If  $\kappa$ is positive,  there will be  no   restriction on the energy density $\rho$ of dust, so that,  
it  starts to increase  from the initial value $\rho_0$ until it  diverges as the scale 
factor of the collapse vanishes. Therefore,  similar to GR,  the collapse ends up with a singularity at the center of the star.

\begin{center}
\begin{figure*}[htbp]
\centering
\includegraphics[width=7.5cm]{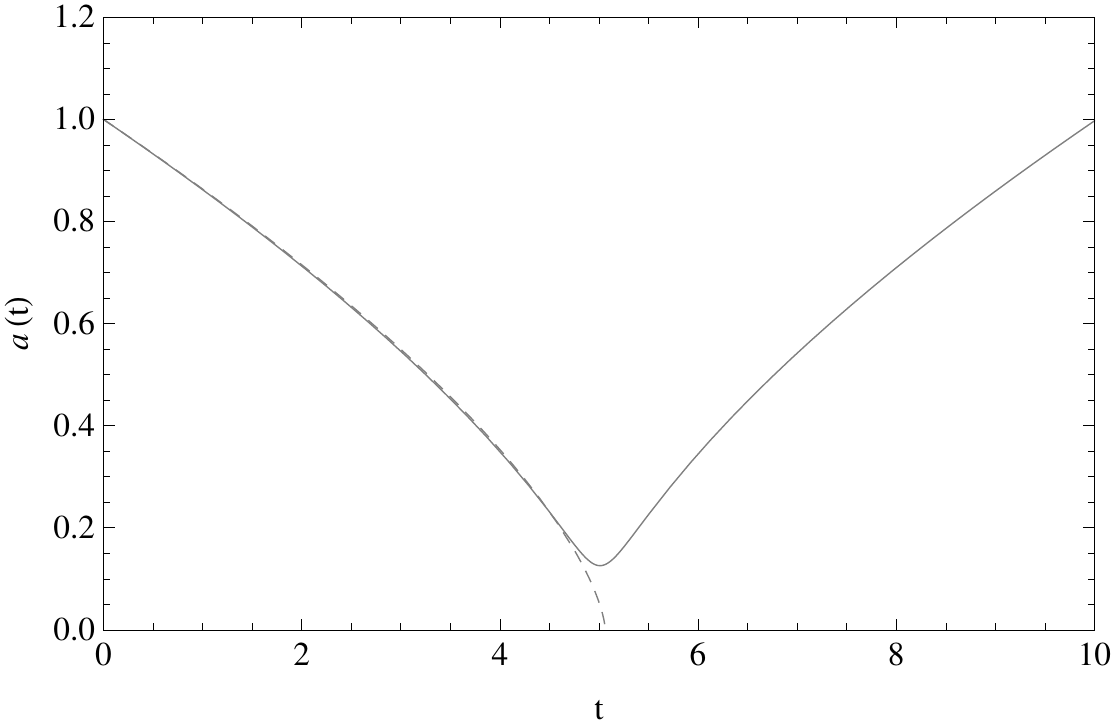}  \quad\quad\quad \includegraphics[width=7.8cm]{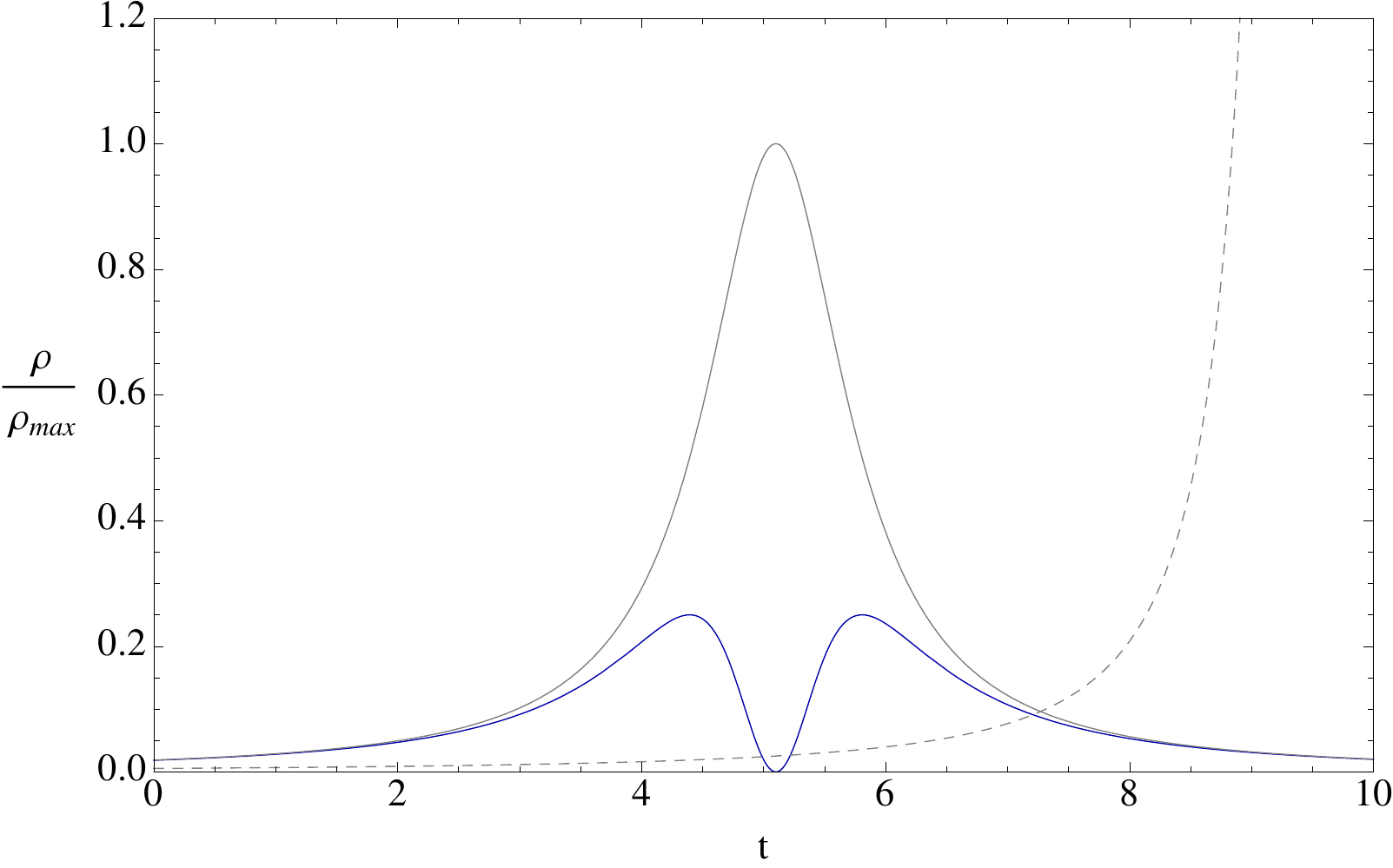}
\caption{{\em Left:}  evolution of scale factor $a(t)$.  
{\em Right:}   evolution of energy density $\rho$  of  dust (gray curve) and the effective density $\rho_{\rm eff}$ of the collapse (blue curve). 
The solid lines  represent  the case in EiBI theory,  while the dashed lines represent  the GR limit. }
\label{ScaleFactor} 
\end{figure*}
\end{center}

The left plot in Fig. \ref{ScaleFactor} shows the evolution of scale factor $a(t)$ of the collapsing cloud  in the interior region. 
It is clear that, the star starts collapsing from an initial scale factor $a_0=a(0)$
and bounces at $a_{\rm B}=(\rho_0/\rho_{\rm max})^{\frac{1}{3}}a_0$. 
In the right plot, the `gray curve'  shows  evolution of the energy density of the  dust, from its initial condition $\rho_0$, until it reaches its maximum  $\rho=\rho_{\rm max}$,
at the bounce. The evolution of the effective energy density $\rho_{\rm eff}$ is also shown by the 
`blue  curve'.  In addition, dashed  curves  represent  the general relativistic
collapse of dust matter which leads to the formation of a singularity at the center of the star.

Consequently,  by expanding Eq.~(\ref{Friedmann-der-formal}) until  second order in $\kappa\rho$, 
we  find the time derivative of the Hubble rate as
\begin{eqnarray}
2\dot{H}\ =\ -\rho\Big(1+\frac{3}{4}\kappa\rho\Big) + {\cal O}\big(|\kappa|^2\big)\ .
\label{Friedmann-der}
\end{eqnarray}
Then, from Eqs.~(\ref{Friedmann}) and (\ref{Friedmann-der}) 
we can define the modified Raychadhuri equation in the herein EiBI theory as 
\begin{eqnarray}
-6\big(\dot{H}+H^2\big) \  =: \ \rho_{\rm eff}+3p_{\rm eff}  \  ,
\end{eqnarray}
so that,  we find an effective pressure for the system as
\begin{eqnarray}
p_{\rm eff}\ \approx \     \frac{3}{8}\kappa \rho^2 \  . 
\label{eq:p_effective}
\end{eqnarray}
Notice that, in the GR  limit,  $\kappa\rho \ll 1$,  we recover the pressureless dust with $p_{\rm eff}=p=0$, and $\dot{H}+H^2=-\rho/6$ (where $\rho_{\rm eff}=\rho$). 
In the bouncing scenario, since $\kappa<0$, 
we get  $p_{\rm eff}= -(3/8) |\kappa| \rho^2=-\rho^2/\rho_{\rm max}<0$.
At the maximum energy ($\rho_{\rm max}=8/3|\kappa|$ at the bounce), pressure reaches its minimum with negative sign as $p_{\rm min}=-8/3|\kappa|=-\rho_{\rm max}$,
which provides a strong repulsive  force near the bounce.
Henceforth, we   will  consider   only  the bouncing scenario with $\kappa<0$.

It is straightforward to show that,  the effective  density (\ref{Friedmann}) and  pressure (\ref{eq:p_effective}) satisfy the (effective) conservation equation: 
\begin{eqnarray}
\dot{\rho}_{\rm eff}+3H(\rho_{\rm eff}+p_{\rm eff})=0.
\label{cons-eff}
\end{eqnarray}
Indeed, by substituting $\rho_{\rm eff}$ and $p_{\rm eff}$ in equation (\ref{cons-eff}), we obtain $(\dot{\rho}+3H\rho)(1-2\rho/\rho_{\rm max})=0$, which leads to 
$\dot{\rho}+3H\rho=0$; this  is  the standard conservation
equation for a dust fluid  which is  satisfied in the herein EiBI  theory (cf. see paragraph under Eq.~(\ref{main-eqn2})).

Let us analyse  the status of the energy conditions  \cite{Hawking:1974}  for the effective matter content, in the interior region of herein collapsing scenario.
For the effective energy density and pressure, we always have  $\rho\geq0$ and $\rho_{\rm eff}\geq0$, so  {\em weak}  energy condition is  always satisfied.
However, in the region   in which   $\rho_{\rm max}/2<\rho<\rho_{\rm max}$,   {\em null} energy condition
($\rho_{\rm eff}  + p_{\rm eff}\geq0$)  is violated.
The {\em dominant}  energy condition  implies  that  $\rho_{\rm eff}  - |p_{\rm eff}|=\rho\geq0$, so,  it is always satisfied.
Moreover, the  {\em strong} energy condition ($\rho_{\rm eff}  + 3p_{\rm eff}\geq0$)
is  violated in the range  $\rho_{\rm max}/4<\rho<\rho_{\rm max}$. 
Note that, this result is consistent with general property of violation of energy conditions  in the context of full EiBI theory \cite{Delsate:2012}. 
Therefore, the singularity resolution in the herein effective scenario of EiBI theory is associated with the violation
of (effective) energy conditions, which may  suggest  that the
EiBI modifications  to our gravitational system   provide  a repulsive force at the very short distances.

\subsection{The exterior black hole structure}
\label{exterior-sol}

Let us consider that $M_0=(4\pi/3)\rho R^3$ (i.e., the total mass of the star), where $\rho=\rho_0(a_0/a)^3$,  and $R(t)=r_ba(t)$   is  
the proper radius from the center  of the cloud 
(with $r_b$ being  the radius of the boundary shell $\Sigma$). 
Note that,  in the interior comoving coordinates, the collapsing boundary surface $\Sigma$ is given  as a free-fall surface.
Now, in terms of the proper radius $R$, we can write   Eq.~(\ref{Friedmann}) as 
\begin{eqnarray}
\dot{R}^2\ =\  \frac{M_0}{4\pi R} + \frac{9\kappa M_0^2}{128\pi^2 R^4}  \cdot
\label{Friedmann2-a}
\end{eqnarray}
For the later conveniences, from now on we will restore the Newton constant $G$ in the above formula, by changing  the mass as $M_0\rightarrow M\equiv M_0/8\pi G$ (bearing in mind that $\kappa\rho$ in Eq.~(\ref{Friedmann}) is dimensionless), and  rewrite Eq.~(\ref{Friedmann2-a}) as
\begin{eqnarray}
\dot{R}^2\ =\  \frac{2GM}{R} + \frac{9\kappa GM^2}{16\pi R^4}  \cdot
\label{Friedmann2}
\end{eqnarray}
By setting $\dot{R}^2=0$ in Eq.~(\ref{Friedmann2}), we can find the location of the bounce, $R_{\rm B}$, as
\begin{eqnarray}
R_{\rm B}\ =\ \left(\frac{9|\kappa|M}{32\pi}\right)^{\frac{1}{3}} \  =\  r_ba_{\rm B}\ .
\label{R_B}
\end{eqnarray}
This indicates that, the physical range for evolution of  the area  radius is $R_{\rm B}<R<R_{0}$ (with $R_0=r_ba_0$).

In order to complete the full space-time geometry, we need to match the homogeneous
interior space-time to a suitable exterior geometry.
From the point of view of an effective scenario for the interior region, it is seen that, 
the `form' of the geometric sector of the theory remains unchanged with respect to GR (see the left hand side of the Friedmann equation~(\ref{Friedmann})), 
while  the matter sector (on the right hand side of Eq.~(\ref{Friedmann})) is replaced by an effective matter content (represented  by an effective density 
$\rho_{\rm eff}$, and an effective pressure $p_{\rm eff}$).
Therefore, it is  reasonable  to employ  the standard matching conditions of GR for the 
(effective) geometric  sector of our model, in order to  investigate  the effects of modified interior  on the emergence   of a possible (effective)  exterior
metric. This  allows the description of the physical consequences that occur at the late time evolution of the collapse in the herein effective scenario of EiBI theory.
We henceforth assume that,  the exterior metric  is static
and satisfies the standard Darmois-Israel  junction conditions at the boundary surface $\Sigma$.
We would then  check
whether this exterior is physically relevant  by imposing the vacuum condition, $T^{ab}=0$,  to the  EiBI equations  of motion  (see Eqs.~(\ref{Eq-Motion1}) and (\ref{Eq-Motion2})).

We consider the most general {\em static} spherically symmetric  metric  that could
match the interior region on the boundary $\Sigma$  \cite{Bruni:2001}:
\begin{equation}
ds^2_{\rm (ext)} = -h^2(R)f(R)d\tau^2 + f^{-1}(R)dR^2 + R^2d\Omega^2 ,
\label{metric-ext}
\end{equation}
 where 
\begin{eqnarray}
f(R)\ =\  1 -  \frac{2Gm(R)}{R}\ \cdot
\end{eqnarray} 
The  Darmois-Israel matching conditions  require  
that the metric  be  continuous (i.e., $R=r_ba$) on $\Sigma$,   
and the extrinsic curvature of   $\Sigma$ be continuous \cite{Stephani-Book,Joshi:2007}.
In order to determine the functions $f(R)$ and $h(R)$ it is convenient to rewrite the interior and exterior  metrics   in null coordinates:
By using $dv=d\tau+dR/[h(R)f(R)]$, the exterior metric (\ref{metric-ext}) reads 
\begin{equation}
ds^2_{\rm (ext)} = -h^2(R)f(R)dv^2 + 2h(R)dvdR + R^2d\Omega^2.~
\label{metric-ext-null}
\end{equation}
For the  interior metric (\ref{metric-interior}) in null coordinates,  we obtain \cite{Bruni:2001, Joshi:2007}
\begin{equation}
ds^2_{\rm (int)} = -t_{,v}^2\big(1- \dot{R}^2\big)dv^2 + 2t_{,v} dvdR +  R^2d\Omega^2\ ,
\label{metric-interior-null}
\end{equation}
where $dt=t_{, v}dv + dR/(\dot{R} - 1)$. 
By comparing two metrics (\ref{metric-ext-null}) and (\ref{metric-interior-null}), we get  $f(R)=1-\dot{R}^2$.
On the other hand, 
since $\Sigma$ now  is a free falling surface in both metrics, we obtain the radial geodesic  equation for the exterior metric 
as $\dot{R}^2=-f+C/h^2(R)$, where $C$ is a constant. The last two equations imply  that $C/h^2(R)=1$ and $h$ is a constant.
Without loss of generality,  we can set  $h=C=1$, and 
\begin{eqnarray}
f(R) \ =\  
1- \frac{2GM}{R} + \frac{9|\kappa| GM^2}{16\pi R^4} \ \cdot
\label{Redshift-final}
 \end{eqnarray}
Now,  the effective mass, $m(R)$, of the exterior geometry can be written as
\begin{eqnarray}
m(R)\ =\ M -  \frac{9|\kappa| M^2}{32\pi R^3}  \  ,
\label{mass-EiBI}
\end{eqnarray}
which is not a constant, and  deviates from  the  Schwarzschild mass $M$    with an additional term $-9|\kappa| M^2/32\pi R^3$.
Note that,  since  $R$  is restricted to  the range $R_{\rm B}<R<R_{0}$,
 the  geometry of exterior metric (\ref{metric-ext}) is always regular.

Let us now investigate   whether this exterior is physically relevant,  
by imposing the EiBI  field  equations  in the absence of  sources.
We have seen that  in  vacuum,   Eq.~(\ref{action-EiBI})  is, in principle, 
locally equivalent to the Einstein-Hilbert action, whose equations of motion possess a static Schwarzschild solution, with $f(R)=1-2GM/R$ 
and the scalar curvature ${\cal R}=0$,  in the exterior region  of the collapse.
However,  the geometry of the  exterior solution (\ref{metric-ext}), 
with the exterior function (\ref{Redshift-final}), and a
  scalar curvature:
\begin{eqnarray}
{\cal R}\ =\   \frac{9|\kappa| GM^2}{\pi R^6} \ ,
\label{Ricci-scalar-3}
\end{eqnarray}
is  different  from  that  of  the Schwarzschild one, thus, 
a static solution is possible only if  $\kappa M\approx 0$.
In the one hand,
the formation of a non-Schwarzschild solution in the exterior region is   a consequence of the fact that,  
in the herein effective  gravitational scenario,  the  pressure governing the collapse evolution  is given by
an  effective  pressure  (\ref{eq:p_effective}), which  includes  a first order correction in $\kappa$, and  does not  vanish  at the boundary surface $\Sigma$  \cite{Sham:2013}
(cf. see also similar effective collapse scenarios which were  provided previously by  brane-world  \cite{Bruni:2001} or quantum gravity \cite{Tavakoli:2014} models).

On the other hand, this departure from  the GR features in vacuum is due to the presence of the  first order contribution of $\kappa$ to the field equations (\ref{Eq-Motion1}) and (\ref{Eq-Motion2}).
More precisely, when the corrections in $\kappa$, up to `first order' is considered, 
the Ricci tensor ${\cal R}_{ab}$ constitutes of  first and  higher order terms  in $\kappa$.
It turns out that,  the second term in equation of motion (\ref{Eq-Motion1}) will be of  {\em second} order in $\kappa$, 
so, it is  negligible in our  first  order approximation. Thus,  
the field equation   (\ref{Eq-Motion1})   
can be written  as\footnote{cf. see appendix \ref{appendix1} for more details.},
\begin{equation}
q_{ab}\ \approx\  g_{ab}+{\cal O}(|\kappa|^2)\ .
\label{Eq-Motion1-b}
\end{equation}
Using this in Eq.~(\ref{Eq-Motion2}), the stress-energy tensor   reduces to   
\begin{eqnarray}
\tilde{T}_{ab}\approx \kappa T_{ab}^{(1)} + {\cal O}(|\kappa|^2), 
\label{energy-ten-final}
\end{eqnarray}
which gives rise to  a  non-Schwarzschild solution, 
with the redshift function  (\ref{Redshift-final}),  for the exterior geometry. 
By using the relations (\ref{Ricci-tensor00})-(\ref{Ricci-tensor33}) we can rewrite the gravitational field equation, to the first order approximation (\ref{Eq-Motion1-b}), as ${\cal R}^a_b-\frac{1}{2}{\cal R}g_b^a\approx 8\pi G \tilde{T}_b^a$, in which  we obtain the effective energy momentum tensor, corresponding to (\ref{energy-ten-final}), as
\begin{eqnarray}
\tilde{T}_b^a \ \propto\  \frac{\kappa M^2}{R^6}+{\cal O}(|\kappa|^2)\ .
\label{energy-ten-final2a}
\end{eqnarray}
In other words, at the effective level, geometry of the exterior region is governed by the Einstein's field equations, which contain a  modified (non-zero) energy-momentum tensor as matter source,  leading consequently  to a non-Schwarzschild solution.

By setting $f(R_h)=0$,  we  can find the location of horizon, $R_h$, 
in the case of a black hole forming at the final state of the collapse.
This gives
\begin{eqnarray}
R_h^4 - \big(2GM\big)R_h^3 + \big(9|\kappa| M^2G/16\pi\big)  =0\ .
\end{eqnarray}
Let us look for  the extremum points of $f(R)$ by solving $f^\prime(R_m)=df/dR|_{R_m}=0$. We obtain
\begin{eqnarray}
R_{m}\ =\  \left(\frac{9|\kappa|M}{8\pi}\right)^{\frac{1}{3}}\ .
\label{R_m}
\end{eqnarray}
By computing  $f^{\prime\prime}(R_m)=d^2f/dR^2|_{R_m}$, it follows that 
$f^{\prime\prime}(R_m)>0$,  thus, $R_m$ is a minimum point.
By comparing (\ref{R_m}) with Eq.~(\ref{R_B}) we find  that $R_{m}=4^{1/3}R_{\rm B}>R_{\rm B}$.

\begin{center}
\begin{figure}[htbp]
\centering
\includegraphics[width=8.5cm]{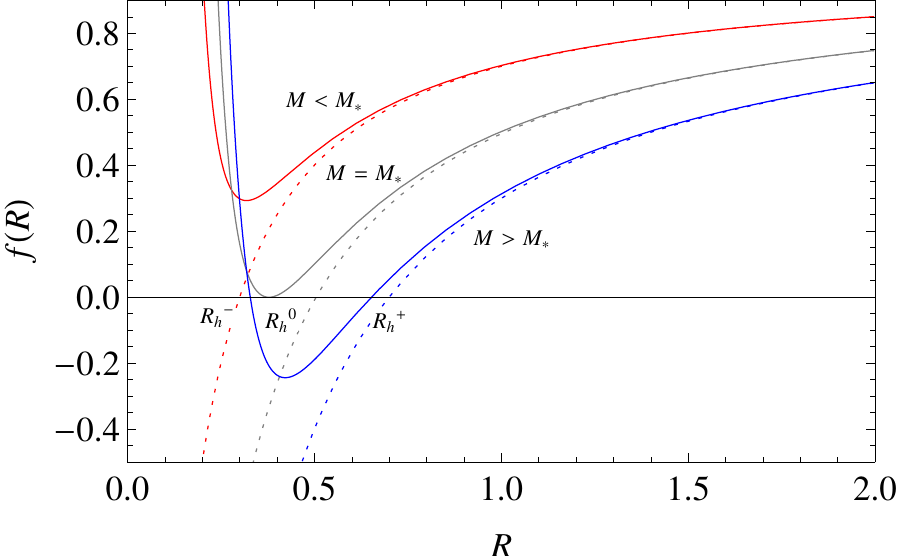}
\caption{\label{RedShift} Solid curves represent the evolution of the redshift function 
(\ref{Redshift-final}), in EiBI theory, for  different values of  mass $M$ of the star. 
The corresponding dotted  curves represent  the  GR  limits.}
\end{figure}
\end{center}

If the value of $f(R)$ at minimum point $R_m$ vanishes, the 
exterior function $f$ has only one root. And moreover, $R_m$ itself  is the root
of $f(R_m)=0$. In other words, $R_m$ is the location of the horizon in this case; $R_h^0=R_m$. Now,
by substituting $R_m$ from (\ref{R_m}) in $f(R_m)=0$ and solving for $M$,  we find a relation for the mass of the star as
\begin{eqnarray}
M_\star\ =\  \left(\frac{|\kappa|}{3\pi G^3}\right)^{\frac{1}{2}}\ .
\label{M-star}
\end{eqnarray}
This analysis indicates that, if the mass of the star is equal to $M_\star$,
the minimum of $f$ vanishes at some point $R_m=R_h^0$. Therefore, the final state  of
the exterior geometry will be a black hole with a double horizon (an extremal black hole). 
The evolution of $f$ in this case is shown by the  `gray curve' in Fig.~\ref{RedShift}. 
If $f(R_m)<0$,   the exterior function $f$ crosses  the horizontal axe, so,  it  will have two roots. 
Similar to the previous analysis,
it can be shown that,  the mass of the  star is  bigger than $M_\star$ in this case. 
Therefore,  the corresponding space-time possess     two horizons:
an  interior horizon,  denoted by  $R_{h}^-$,  and an  exterior one, 
denoted by  $R_{h}^+$ (see `blue curve' in Fig.~\ref{RedShift}).
Finally, for the case  $M<M_\star$,  the exterior function $f$ is always  positive at its minimum point, $f(R_m)>0$,
therefore,  no horizon would form during the collapse   (see `red curve' in Fig.~\ref{RedShift}).

For the  collapsing star of dust matter  with the mass $M$, 
the energy density of the interior region  implies that
$M=(4\pi \rho/3) R^3$. This gives a lower scale limit  $R_{\rm B}$ at which collapsing star bounces:
\begin{eqnarray}
R_{\rm B}\ \approx \  \left(\frac{3}{4\pi\xi}\frac{M}{m_{\rm Pl}}\right)^{\frac{1}{3}}\ell_{\rm Pl}\ ,
\label{bounce-mass}
\end{eqnarray}
where,  $\ell_{\rm Pl}=\sqrt{\hbar G/c^3}\sim 1.67\times10^{-35}~m$ is the Planck length,  $m_{\rm Pl}=\rho_{\rm Pl}\ell^3_{\rm Pl}=\sqrt{\hbar c/G}\sim 2.18\times10^{-8}~kg$ is the Planck mass, and $\rho_{\rm Pl}=c^5/\hbar G^2\sim 5.15\times10^{96}~kg\cdot m^{-3}$ is the Planck density.
Here we have assumed  that,   the energy density of the collapsed  star  at the bounce is 
$\rho_{\rm max}\approx \xi\rho_{\rm Pl}$, where $\xi$ is a positive constant smaller than one \footnote{For example, in loop quantum cosmology, bounce occurs at $\xi\approx 0.41$ \cite{APS:2006,Ashtekar-Singh}.}.
Since the initial  stellar mass $M$  is much bigger than the Planck mass,   the bounce might  happen at some orders of magnitude much larger than the Planck volume (cf. see Eq.~(\ref{bounce-mass})). However, in a realistic situation, the stellar mass $M$ does  not remain  constant necessarily and may reduce due to quantum evaporation phenomena during the collapse (cf. see for example \cite{Hayward:2006}).

At the bounce  the energy density of the collapsed star  has its upper bound, so that for  the choice $\xi \approx 1$, from $\rho_{\rm max}=8/3|\kappa| \approx \rho_{\rm Pl}$  we find a lower limit  for the EiBI coupling parameter as\footnote{Notice that,  $\kappa\rho$ in the Friedmann Eq.~(\ref{Friedmann}) is dimensionless.}
\begin{eqnarray}
|\kappa|\ \approx\ 5.1\times10^{-97}~kg^{-1}\cdot m^3~.
\label{kappa-min}
\end{eqnarray}
By replacing the minimum  value of $|\kappa|$ from Eq.~(\ref{kappa-min}) in Eq.~(\ref{M-star}) (when restoring $c$ therein by $G\rightarrow G/c^2$),  the   mass threshold $M_\star$ becomes
\begin{eqnarray}
M_\star\ \approx\  \ 
0.53~ m_{\rm Pl}\ .
\label{M-star2}
\end{eqnarray}
Therefore, the mass (\ref{M-star2}) of the final black hole is smaller than the Planck mass, which represents  the smallest black hole that can exist in the herein EiBI model. 
This threshold mass is comparable with the minimum  mass of a (micro) black hole predicted in GR. Indeed, at Planck scale, the Compton wavelength limitation on the minimum size of the Schwarzschild radius  in GR  implies that  the smallest mass of any micro black hole should approximately be the Planck mass.

A possible scenario for the outcome of our  stellar collapse, with an initial mass $M$ much  bigger than the  mass threshold $M_\star$, is  that a non-singular  black hole  will form with an inner and an outer horizons. Then, as collapse proceeds, a quantum evaporation mechanism occurs during which  the black hole mass $M$ reduces from its initial value until it reaches  the threshold of  mass  $M_\star$ \cite{Hayward:2006}. Hereafter, an extremal, non-radiating black hole  with a constant  valued mass $M_\star$ remains. 
The remaining  black hole with the final mass $M_\star$ may also continue collapsing   to a Planck star with the size very small compared to the original star \cite{Rovelli:2013}.  From  the radius of the horizon $R_m=R_h^0=4^{1/3}R_{\rm B}$ of such black hole  (with mass (\ref{M-star2}) and $\xi\approx1$),  we obtain\footnote{See Refs.~\cite{Rovelli:2013, Hossenfelder:2010} for a similar estimate of horizon size.}
\begin{eqnarray}
R_h^0\ \approx\  \left(\frac{3}{\pi}\frac{M_\star}{m_{\rm Pl}}\right)^{\frac{1}{3}}\ell_{\rm Pl} \ \approx \
 0.79~\ell_{\rm Pl} \ ,
\label{R0-min}
\end{eqnarray}
which represents the radius of the {\em smallest}  possible black hole that can exist in the herein EiBI theory of gravity.

\section{Geodesic behaviours}

\begin{center}
\begin{figure}[htbp]
\centering
\includegraphics[width=8.6cm]{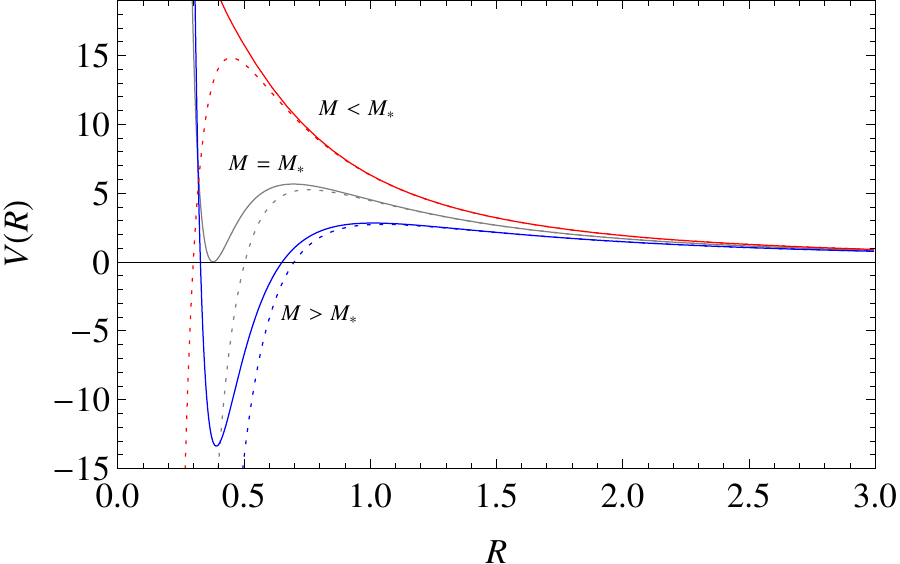}
\caption{Evolution of the potential (\ref{potential-1}) for  different values of the mass of the star. 
Solid curves represent the potential in EiBI theory, 
while dotted  curves correspond to  GR limits.}
\label{Potential} 
\end{figure}
\end{center}

In order to study the  behaviour of the null rays in the exterior region of the herein collapsing system, 
we consider the null geodesics equation for the exterior geometry (\ref{metric-ext}) which
can be written as
$E^2 =  ( dR/d\gamma)^2+2V(R)$.
Here,  $E$ is the energy of  massless particles,   
$\gamma$ is an affine parameter, and  $V(R)$ is the effective potential  given by  
\begin{equation}
V(R)\ =\ \frac{L^2}{2R^2}\left(1- \frac{2Gm(R)}{R}\right)\ .
\label{potential-1}
\end{equation}
Moreover,  $L=R^2(d\varphi/d\gamma)$ is  the angular momentum of massless particles (we may interpret $\hbar L$ 
as the angular momentum of a photon moving on the exterior geometry) \cite{Wald:1984}.
The crucial new feature provided by EiBI correction  term in
(\ref{mass-EiBI})    is that,  we obtain  an additional centrifugal barrier term, 
$9L^2|\kappa|GM^2/(32\pi R^6)$,   in Eq.~(\ref{potential-1}),   which dominates over the standard Schwarzschild terms at small $R$.
Fig.  \ref{Potential}  shows  the behaviour of the potential (\ref{potential-1}) in the herein EiBI  modified  regime, 
that, unlike the classical Schwarzschild geometry,  depends on the values of $M$.

For a massless particle (or a photon) with energy $E$, moving along  the potential (\ref{potential-1}),  
there is  a  `turning point'   at $V(R)=E^2/2$, from which  it will begin to move  in the opposite  direction. 
If the potential is flat,  i.e., $dV/dR=0$, the photon
may simply move in a circular orbit at radius $R_c=const.$, which is given by the solution of
\begin{equation}
R_c^4 - \big(3GM\big)R_c^3  + \big(27|\kappa| GM^2/16\pi\big)=0\ .
\label{quartic-2}
\end{equation}
For the case $M<M_\star$, the equation above has no root, thus,
there is  no stable circular orbit  (see `red curve' in Fig.~\ref{Potential}).
A sufficiently energetic photon 
coming  from infinity,   slows down gradually  (with decreasing  $dR/d\gamma$) 
until it reaches the turning point $V(R)=E^2/2$, 
 then, it  starts  moving  back to infinity.
For the cases $M\geq M_\star$,  Eq.~(\ref{quartic-2}) has two roots which correspond to  two extremal points for the potential:
there exists a radius $R_{c}^-$ at which the potential is  minimum, i.e.,  $V(R_{c}^-)=V_{\rm min}$, 
and a radius $R_{c}^+$ at which the potential has a  maximum,  $V(R_{c}^+)=V_{\rm max}$.
Therefore, circular orbits with the radius $R_{c}^-$ are stable and those with the radius $R_{c}^+$ are unstable. 
For the case $M=M_\star$, the stable circular orbits are located at the horizon of the (extremal) black hole, 
that is,  $R_{c}^-=R_h^0=(9|\kappa| M_\star/8\pi)^{\frac{1}{3}}$. This means that,  
any  photon   (with an energy  higher  than the energy of the barrier; $E^2>2V_{\rm max}$) 
coming  from infinity, would  be trapped  on the black hole  horizon  and  starts   a (stable) circular orbiting   on the surface of the   horizon.
In the case $M>M_\star$, stable circular orbits occur at $R_{c}^-$, where  $R_h^-<R_{c}^-<R_h^+$,  
thus,    any    sufficiently energetic 
photon   coming   from infinity will be trapped somewhere between the inner and outer horizons. 
In the GR limit,  there is only one {\em unstable} circular orbit for any  photon; 
a small perturbation will push back  the photon to  infinity, or,  it will  fall inside the black hole horizon until it reaches the singularity at $R=0$.

\section{Conclusion and discussions}

A spherically symmetric model for gravitational collapse of a homogeneous dust ball 
was considered in the framework of the
Eddington-inspired Born-Infeld (EiBI) theory of gravity. We obtained an approximate effective  expression for the Friedmann 
equation governing the evolution equations of the collapse. It was shown that, the energy density of the collapsing matter 
has an upper bound, $\rho_{\rm max}=8/3|\kappa|$, for   negative values of the  EiBI coupling parameter $\kappa$.  
When the energy density of the matter
reaches  this  maximum value, the collapse hits a bounce  before  
it approaches  the center of the star, where  a GR   singularity is located.
Furthermore, we  studied the energy conditions, where an effective matter  density and  pressure raised due to  the modified equations of motion. We have shown that,
the (effective) `null' and `strong' energy conditions are not satisfied  in the regions  close to the classical singularity (cf.  Ref. \cite{Delsate:2012}). 
This indicates that  the singularity resolution in the herein effective scenario of EiBI gravity, 
is associated with the violation of energy conditions at the late time stages of  gravitational collapse.

We  investigated  different  possibilities   for  the  exterior geometry  
as the final state of the gravitational collapse in the herein effective  EiBI  theory.
Our analysis provided  three scenarios  for the horizon formation depending on the initial conditions  of the collapse. 
In particular, we obtained a  mass threshold, $M_\star=(|\kappa|/3\pi G^3)^{\frac{1}{2}}$, 
 for the star, below which no horizon would form during the collapse.
When the initial mass is equal and bigger than the  critical mass, $M\geq M_\star$, one and two horizons would form, respectively. 
By   assuming  that  the  maximum energy density of the collapsed star at the bounce is the Planck density,  we  found  a {\em minimum} value  for the EiBI coupling  parameter  as $|\kappa|=5.1\times10^{-97}~kg^{-1}\cdot m^3$.
This small value of $\kappa$ is compatible with our approximations and makes the theory indistinguishable of GR in the low energy limit. This further estimates that  the mass of the smallest possible black hole in the Universe  is  approximately $0.53~m_{\rm Pl}$.
We concluded our work by analyzing   the  geodesic behaviour  of 
any massless particle propagating on the geometry of the resulting  black holes. 
We have shown that, in the case of the extremal black hole (with $M=M_\star$), a massless particle,  
coming in from infinity,  will be trapped  on   the horizon of the black hole and starts a (stable) circular orbit around it.  Moreover, 
for  $M>M_\star$, any sufficiently energetic massless particle near   the exterior horizon will   fall   inside the  horizon and starts circulating 
at  the minimum  of the potential, somewhere between the inner and outer  horizons.

In the   herein effective EiBI  scenario, up to the `first'  order corrections in $\kappa$ to the  field equations in the interior region (cf. see Eq.~(\ref{Friedmann})), 
the exterior (regular) black hole geometry  
was obtained  due to the standard Israel-Darmois matching%
\footnote{It should be noticed that, as  long  as an {\em effective}  theory is in hand,  in which the form of the equations of motion looks  like  those  in GR, 
application  of the standard general relativistic  matching conditions is   relevant.}
at  the boundary of two regions.
In this  approximation, 
in order to the exterior solution (\ref{metric-ext-null}) 
with redshift function (\ref{Redshift-final}) 
be consistent with the equations of motion (\ref{Eq-Motion1}) and (\ref{Eq-Motion2}),
the (effective) energy-momentum tensor $\tilde{T}^{ab}$ 
in the outer region should have a linear dependence 
on $\kappa$ (cf. see Eqs.~(\ref{energy-ten-final}) and (\ref{energy-ten-final2a})). 
In other words, the effects of the EiBI modifications to the interior region are carried out to the
exterior region, due to matching at the boundary surface, 
which induce  an effective energy-momentum tensor of the form 
$\tilde{T}^{b}_a\approx\kappa M/R^6 + {\cal O}(|\kappa|^2)$. 
This  induced matter 
implies the  non-Schwarzschild geometry (\ref{Redshift-final}) for the exterior.

\begin{acknowledgements}

YT thanks Bonyad-e-Melli Nokhbegan of Iran (INEF) for financial supports.
He also acknowledges Brazilian agencies   FAPES and CAPES  for partial  financial supports. 
CE-R is supported by a  CNPq  fellowship  (Brazil).  
JCF  thanks    CNPq  and FAPES  for partial financial support.  
This work  was also supported  by   the project  CERN/FP/123609/2011.

\end{acknowledgements}

\appendix

\section{Higher order corrections in $\kappa$ to  field equations}
\label{appendix1}
 
For the exterior metric (\ref{metric-ext}) we have calculated the exterior function (\ref{Redshift-final})  to the first order corrections in $\kappa$:
\begin{eqnarray}
f(R)=1-\frac{2GM}{R} + \frac{9|\kappa|GM^2}{16\pi R^4} +  \mathcal{O}(|\kappa|^2)\ .
\label{Redshift}
\end{eqnarray}
For this metric, the  Ricci tensor components ${\cal R}_{ab}$ are given by
\begin{eqnarray}
\mathcal{R}_{00} &=& \frac{12G\beta}{R^{10}}\big(-R^4+2GR^3\alpha+2G\beta\big) +  \mathcal{O}(|\kappa|^2) \nonumber \\
 &\ =:&  |\kappa|M^2 F_0(R)+ \mathcal{O}(|\kappa|^2)   , \label{Ricci-tensor00} \\
\mathcal{R}_{11} &=& \frac{12G\beta}{R^6-2GR^5\alpha-2GR^2\beta}
+ \mathcal{O}(|\kappa|^2) \nonumber \\
&\ =:& \  |\kappa|M^2 F_1(R)+ \mathcal{O}(|\kappa|^2)  , \label{Ricci-tensor11} \\
\mathcal{R}_{22} &=& -\frac{6G\beta}{R^4} + \mathcal{O}(|\kappa|^2)
\nonumber \\ &\ =:&\  |\kappa|M^2 F_2(R) + \mathcal{O}(|\kappa|^2)  , \label{Ricci-tensor22} \\
\mathcal{R}_{33} &=& - \frac{6G\beta\sin^2\theta}{R^4} + \mathcal{O}(|\kappa|^2) \nonumber \\
 &\ =:&   |\kappa|M^2 F_3(R) + \mathcal{O}(|\kappa|^2)   ,\label{Ricci-tensor33}
\end{eqnarray}
where $F_a$s  are some functions of physical  radius $R$.
Notice that, we have defined,
$\beta\equiv\frac{9|\kappa|M^2}{32\pi}$.
By substituting the components  of the Ricci tensors (\ref{Ricci-tensor00})-(\ref{Ricci-tensor33}) in the equation of motion (\ref{Eq-Motion1}), up to the first order approximation in $\kappa$,  we obtain
\begin{eqnarray}
q_{ab}=g_{ab} + \mathcal{O}(|\kappa|^2)\ .
\label{motion-EiBI1}
\end{eqnarray}
Therefore, the second term in equation of motion (\ref{Eq-Motion1}), is at least to the {\em second} order in $\kappa$, which is  negligible in our (first order) approximation.
Notice that, here we have defined the Ricci tensor ${\cal R}_{ab}$  in terms of the (exterior) metric $g_{ab}$, but it should have been defined in terms  of  $q_{ab}$. However,  in the first order approximation since $q_{ab}\approx g_{ab}$, the choice of the metric to define ${\cal R}_{ab}$ is the same for both metrics (that is ${\cal R}_{ab}[q]\approx {\cal R}_{ab}[g]$), thus, our result is consistent with (\ref{Eq-Motion1}). 
In order to have this approximation ($q_{ab}\approx g_{ab}$) consistent with the other equation of motion, Eq.~(\ref{Eq-Motion2}), 
the (exterior) stress-energy tensor $T^{ab}$ 
should be zero or have a linear dependance on $\kappa$:  
\begin{eqnarray}
T^{ab}_{\rm (ext)}\ &=& \ \frac{1}{\kappa\sqrt{g}} \big(\lambda\sqrt{g}g^{ab} - \sqrt{q}q^{ab}\big) \nonumber \\
&\approx& \  \kappa~T^{ab}_{(1)} + {\cal O}(|\kappa|^2)\ =: \ \tilde{T}^{ab}
\label{energy-momentum-1}
\end{eqnarray}
Here $T^{ab}_{(1)}$ is a function of $R$ (cf. see Eq.~(\ref{energy-ten-final2a})). 
Since, in the absence of sources in the exterior region, the   EiBI  equations of motion should reduce to the vacuum Einstein field equation, hence,  the exterior geometry should  be the Schwarzschild one.  However, the exterior (\ref{Redshift}) (induced by matching at the boundary of collapsing cloud)  in the first order corrections in $\kappa$, is different from Schwarzschild. This implies an (induced) non-zero energy-momentum tensor (\ref{energy-momentum-1}) for the exterior region.

For the sake of completeness, let us discuss, 
following the same method we presented in  paragraph above, the case
where higher order correction terms in $\kappa$ is present in the
EiBI equations of motion (\ref{Eq-Motion1}) and (\ref{Eq-Motion2}).
By considering the second  order corrections in $\kappa$ in the 
Friedmann equation (\ref{Friedmann-formal}), we obtain
\begin{eqnarray}
3\frac{\dot{a}^2}{a^2}\    \  =\    \rho\Big(1+ \frac{3}{8}\kappa\rho\Big)  -\frac{27}{16}\kappa^2\rho^3 + {\cal O}\big(|\kappa|^3\big) \ .
\label{Friedmann-extra}
\end{eqnarray}
For the new  Friedmann equation (\ref{Friedmann-extra}) in the interior region, by imposing the Israel-Darmois  matching conditions  
(as in section \ref{exterior-sol}) at the boundary of two regions, we obtain the exterior metric (\ref{metric-ext}) for which the redshift function is  given by 
\begin{eqnarray}
f(R) &=&1-\frac{2GM}{R} + \frac{9G}{16\pi}\frac{|\kappa|M^2}{R^4} \nonumber \\
&& \quad +  \frac{81G}{32\pi^2}\frac{|\kappa|^2 M^3}{R^7} + \mathcal{O}(|\kappa|^3)\ .
\label{Redshift2}
\end{eqnarray}
Then, we need to compute the Ricci tensor ${\cal R}_{ab}[q]$, in the presence of the new exterior function 
(\ref{Redshift2}) including the second  order  term  of  $\kappa$, in order  to investigate  the consistency  of this  effective scenario.
As before, let us compute ${\cal R}_{ab}$ for the metric component $g_{ab}$, given by Eq.~(\ref{metric-ext}) with the redshift (\ref{Redshift2}). We obtain
\begin{eqnarray}
\mathcal{R}_{00} &&=   \frac{6G}{R^{16}}\big(2R^3\beta+7\lambda\big)\big(-R^7+2GR^6\alpha
\nonumber \\
&& \quad \quad \quad +2GR^3\beta +2G\lambda\big) +  \mathcal{O}(|\kappa|^3) \nonumber \\
&&=:   |\kappa|M^2\tilde{F}_0(R) + |\kappa|^2M^3\tilde{G}_0(R)+  \mathcal{O}(|\kappa|^3), \quad \quad \quad  \label{Ricci-tensor2-00}
\end{eqnarray}
\begin{eqnarray}
\mathcal{R}_{11} &&=\   \frac{6G(2R^3\beta+7\lambda)}{R^9-2GR^8\alpha-2GR^5\beta -2GR^2\lambda} +  \mathcal{O}(|\kappa|^3) \nonumber \\
&&=:  \  |\kappa|M^2\tilde{F}_1(R) + |\kappa|^2M^3\tilde{G}_1(R) +  \mathcal{O}(|\kappa|^3),  \quad \quad  \quad  \label{Ricci-tensor2-11}
\end{eqnarray}
\begin{eqnarray}
\mathcal{R}_{22} &&= \  - \frac{6G}{R^{7}}\big(R^3\beta +2\lambda\big)+  \mathcal{O}(|\kappa|^3) \nonumber \\
&&=:\   |\kappa|M^2\tilde{F}_2(R) + |\kappa|^2M^3\tilde{G}_2(R) +  \mathcal{O}(|\kappa|^3), \quad  \quad \quad \label{Ricci-tensor2-22}
\end{eqnarray}
\begin{eqnarray}
\mathcal{R}_{33} &&= \   -\frac{6G}{R^{7}}\big(R^3\beta +2\lambda\big)\sin^2\theta
+  \mathcal{O}(|\kappa|^3) \nonumber \\
&&=: \    |\kappa|M^2\tilde{F}_3(R) + |\kappa|^2M^3\tilde{G}_3(R) +  \mathcal{O}(|\kappa|^3) , \quad\quad \quad \quad \label{Ricci-tensor2-33}
\end{eqnarray}
where we have used a new definition: 
$\lambda\equiv\frac{81|\kappa|^2M^3}{64\pi^2}$.
By substituting Eqs.~(\ref{Ricci-tensor2-00})-(\ref{Ricci-tensor2-33}) in equation of the field (\ref{Eq-Motion1}), up  to the second order corrections in $\kappa$, we find:
\begin{eqnarray}
q_{ab}\ &=&\ g_{ab} + \kappa\big[\kappa M^2 \tilde{F}_a(R)+\kappa^2 M^3 \tilde{G}_a(R)\big] \nonumber \\
&=&\ g_{ab} + \kappa^2 \big(M^2 \tilde{F}_a(R)\big) + \mathcal{O}(|\kappa|^3), \quad \quad
\label{motion-EiBI1b}
\end{eqnarray}
where $\tilde{F}_a$s and $\tilde{G}_a$s are some functions of $R$.

The above  analysis indicates  that, by considering the  second  or higher order corrections  in 
$\kappa$,  equation of motion (\ref{Eq-Motion1}) (or,  Eq.~(\ref{motion-EiBI1b}) for {\em second} order approximation)  also contains terms up to that order,  in addition to $g_{ab}$. 
Thus, Eq.~(\ref{Eq-Motion1}) {\em cannot}  be   approximated as  $q_{ab}\approx g_{ab}$, so that, 
our analysis of Ricci tensor, Eqs.~(\ref{Ricci-tensor2-00})-(\ref{Ricci-tensor2-33}),  using  the exterior metric $g_{ab}$, given by (\ref{Redshift2}),  is not relevant anymore. Therefore, an exact analysis for computing the Ricci tensor 
in terms of the metric components of $q_{ab}$ is required. 
Nevertheless, our knowledge from first order approximation indicates that,
the exterior region in this case,  is induced by  an effective energy-momentum tensor including
second or higher order  corrections in $\kappa$.


\end{document}